\documentclass[12pt]{article}
\usepackage{amssymb,amsmath,graphicx,latexsym,cite}
\begin{document}
\title{\textbf{Soft asymptotics with mass gap}} 
\author{B.~Holdom\thanks{bob.holdom@utoronto.ca}\\
\emph{\small Department of Physics, University of Toronto}\\[-1ex]
\emph{\small Toronto ON Canada M5S1A7}}
\date{}
\maketitle
\begin{abstract}
From the operator product expansion the gluon condensate controls a certain power law correction to the ultraviolet behavior of the gauge theory. This is reflected by the asymptotic behavior of the effective gluon mass function as determined by its Schwinger-Dyson equation. We show that the current state of the art determination of the gluon mass function by Binosi, Ibanez and Papavassiliou points to a vanishing gluon condensate. If this is correct then the vacuum energy also vanishes in massless QCD. This result can be interpreted as a statement about a softness in the ultraviolet behavior and the consistency of this behavior with a mass gap.
\end{abstract}

Schwinger-Dyson (SD) equations provide a useful tool for the study of dynamical symmetry breaking by providing information about the momentum dependent dynamical mass functions. The existence of these mass functions signals a mass gap, a distortion of the theory in the infrared. The mass function also specifies power law corrections to the asymptotic ultraviolet behavior of the theory. These corrections in turn are related via the operator product expansion to condensates, vacuum expectation values of local operators. For example in massless QCD the dynamical quark mass function solution of the SD equation has an asymptotic behavior that points to a quark condensate appearing in the operator product expansion of two quark fields. The condensate merely encodes a particular effect that the mass gap has on the ultraviolet behavior of the theory.

In contrast to the quark condensate the gluon condensate does not break a symmetry of the theory, and the result is that the perturbative contribution is sensitive to any dimensionful regulator (a UV cutoff). Such an additional explicit breaking of scale invariance can be avoided with a scale invariant regulator, e.g.~dimensional regularization, and in this case the perturbative contribution vanishes at any finite order.\footnote{A related example is the vacuum energy in a free massless field theory, which vanishes by Lorentz and scale invariance. Spurious contributions arise unless a scale invariant regulator is chosen. In the same way we avoid spurious contributions to the gluon condensate.} But even with the choice of a scale invariant regulator, a resummation of a class of diagrams via the renormalization group leads to the infrared Landau pole at a scale $\Lambda$ defined by
\begin{equation}
\ln(\frac{\Lambda}{\mu})=\int_{g(\mu)}^\infty\frac{dx}{\beta(x)}
.\label{e1}\end{equation}
This dimensional transmutation is taken as a signal of a non-vanishing gluon condensate of order $\Lambda^4$. It is also argued that the gluon condensate is needed to cancel or correct particular ambiguities in perturbation theory (renormalons etc.) that are also related to the Landau pole. All this assumes that the Landau pole is more than just an artifact of resummed perturbation theory. Indeed a variety of approaches \cite{a1,Dokshitzer:1995qm,Smekal:1997is,Shirkov:1997wi,Aguilar:2002tc,Fischer:2006ub,Prosperi:2006hx,Aguilar:2009nf,Brodsky:2010ur,Courtoy:2013qca} indicate that the Landau pole does not survive nonperturbative effects that cause the coupling strength to saturate at a finite value in the infrared. Then the $\Lambda$ determined by (\ref{e1}) vanishes. Lattice studies \cite{Bogolubsky:2007ud,Bogolubsky:2009dc,Oliveira:2010xc,Aguilar:2008xm,Aguilar:2010gm,Ayala:2012pb} have verified the associated damping of gluonic fluctuations in the infrared as described by an effective gluon mass function. This is the view we adopt here, in which case a different approach to the gluon condensate is needed. 

In the same way as for the quark condensate, we may view the gluon condensate as just encoding a particular correction to the ultraviolet behavior due to the presence of the mass gap. We compare the operator product expansions for the quark and gluon mass functions ($\Sigma(p^2)$ and $m^2(p^2)$ respectively) \cite{Larsson:1984yv,Lavelle:1991ve}.
\begin{eqnarray}
\lim_{-p^2\rightarrow\infty}\Sigma(p^2)&=&c_1(p/\mu)m_\psi(\mu)+\frac{c_2(p/\mu)\langle\overline{\psi}\psi\rangle_\mu}{p^2}+...\label{e6}\\
\lim_{\scriptsize\shortstack[c]{$-p^2\rightarrow\infty$\\$m_\psi\rightarrow0$}}p^2m^2(p^2)&=&a_1(p/\mu)\langle G_{\alpha\beta}G^{\alpha\beta}\rangle_\mu+\frac{a_2(p/\mu)\langle\overline{\psi}\psi\rangle^2_\mu}{p^2}+...
\label{e7}\end{eqnarray}
Shown in these expansions are the leading gauge invariant terms in the asymptotic behavior of the mass functions. We comment on possible gauge dependent contributions below. The two condensates are purely nonperturbative and each has a renormalization scale dependence. We see the sense in which the gluon condensate is the analog of $m_\psi$ rather than $\langle\overline{\psi}\psi\rangle$, since $m_\psi$ and the gluon condensate govern the leading term in the respective OPE. Once the bare quark mass is set to zero then $m_\psi(\mu)=0$ remains consistent with the SD result for $\Sigma(p^2)$. We note that the quark condensate that is generated contributes only to the subleading terms in both expansions.

The similarity of the role of the leading terms of the OPEs also shows up in the trace of the energy-momentum tensor
\begin{equation}
\Theta_\alpha^\alpha=\frac{\beta}{2g}G_{\alpha\beta}G^{\alpha\beta}+m_\psi(1+\gamma_m)\overline{\psi}\psi
.\label{e8}\end{equation}
This operator statement gives a relation between the vacuum expectation values $\langle\Theta_\alpha^\alpha\rangle$, $\langle G_{\alpha\beta}G^{\alpha\beta}\rangle_\mu$ and $\langle\overline{\psi}\psi\rangle_\mu$, all of which are taken to vanish in perturbation theory. $\langle\Theta_\alpha^\alpha\rangle$ is a physical quantity independent of $\mu$ since it is four times the vacuum energy by Lorentz invariance. The $\mu$ dependence of the other two condensates must then be cancelled by the $\mu$ dependence of their coefficients in (\ref{e8}). The point is that for $\langle\Theta_\alpha^\alpha\rangle$ to be non-vanishing one or both of the leading terms in the OPEs (\ref{e6}) and (\ref{e7}) (that is $m_\psi(\mu)$ and/or $\langle G_{\alpha\beta}G^{\alpha\beta}\rangle_\mu$) needs to be present.

This property of the OPEs a statement about the asymptotic behavior of the fundamental fields and so the question of whether vacuum energy vanishes $\langle\Theta_\alpha^\alpha\rangle=0$ becomes a question of whether a certain ultraviolet boundary condition is satisfied by the theory. Thus for massless QCD ($m_\psi=0$) it becomes a fundamental question as to whether the leading term of the gluon mass OPE is or is not generated by nonperturbative effects that necessarily go beyond chiral symmetry breaking. This highlights the importance of the gluon mass SD equation, which can in principle be used to determine the asymptotic behavior of the gluon mass function. In other words the SD approach could tell us whether vanishing vacuum energy is compatible with a mass gap.

The SD approach provides a nonperturbative framework (along with the lattice) to define the propagators of the fundamental fields at all momenta. The behaviors of gluon and ghost propagators that are emerging are deepening our understanding of confinement and the mass gap. We note though that any SD analysis involves a truncation of the complete SD equations and this introduces uncertainties, especially in the precise shape of the mass functions at low momentum. For our purposes we can have more confidence in the SD results for the gross features of the asymptotic behavior of the gluon mass function. For example the fact that the asymptotic behavior of the quark mass function in massless QCD is consistent with $m_\psi(\mu)=0$ in (\ref{e6}) is a robust result of the SD analysis.

Significant progress towards obtaining a more accurate SD equation for the gluon mass has been made \cite{Binosi:2012sj}. The full SD kernel, which has both one and two loop parts in terms of dressed quantities, has been reduced to a manageable form with the help of the pinch technique \cite{Binosi:2009qm} and the background field method. Ward identities are maintained to reflect the fact that the gluon mass function should not explicitly break gauge symmetries. This is able to sufficiently specify modified vertices $\Gamma\rightarrow\Gamma'=\Gamma_m+V$ involving ``pole vertices'' $V$. The whole analysis takes place in Landau gauge.

The following Euclidean space integral equation for the gluon mass function is obtained \cite{Binosi:2012sj}.
\begin{equation}
m^2(q^2)=-\frac{4\pi\alpha_sC_A}{1+G(q^2)}\frac{1}{q^2}\int \frac{d^4k}{(2\pi)^4} m^2(k^2)\Delta_\rho^\mu(k)\Delta^{\nu\rho}(k+q){\cal K}_{\mu\nu}(k,q)
\label{e2}\end{equation}
\begin{eqnarray}
{\cal K}_{\mu\nu}(k,q)&=&[(k+q)^2-k^2]\{1-[Y(k+q)+Y(k)]\}g_{\mu\nu}\nonumber\\
&-&[Y(k+q)-Y(k)](q^2g_{\mu\nu}-2q_\nu q_\nu)
\end{eqnarray}
The gluon propagator is
\begin{equation}
\Delta_{\mu\nu}^{ab}(q)=\delta^{ab}\Delta_{\mu\nu}(q)=\delta^{ab}(g_{\mu\nu}-q_\mu q_\nu/q^2)\Delta(q^2)
.\end{equation}
The factor $1/(1+G(q^2))$ is identified with $q^2D(q^2)$ where $D(q^2)$ is the ghost propagator, apparently to good approximation. The quantity $Y(k)$ is a one loop sub-diagram in the two loop contribution to the kernel. If evaluated using tree level propagators and vertices it is \cite{Binosi:2012sj}
\begin{equation}
Y(k^2)=-\frac{\alpha_s C_A}{4\pi}\frac{15}{16}\log\frac{k^2}{\mu^2}
.\label{e5}\end{equation}
This introduces a renormalization scale dependence in the SD equation.

The propagators $\Delta(q^2)$ and $D(q^2)$ are obtained from a fit to lattice data. Since the lattice analysis uses a renormalization scale of $4.3$ GeV, this is the choice adopted for $\mu$ \cite{Binosi:2012sj}. The result is an integral equation which is linear in $m^2(q^2)$. The latter is obtained numerically and normalized to agree with the lattice gluon propagator at $q^2=0$.

The authors in  \cite{Aguilar:2013hoa} extend these results to the unquenched case by incorporating the quark loop contribution, where the quark propagator used is obtained from the quark SD equation. They start with the gluon mass solution in the quenched case, where lattice results for the quenched propagators are used in (\ref{e2}), and then by incorporating the quark effects via an iterative procedure they obtain a prediction for the modified gluon propagator in the $n_f=2$ case. The result agrees very well with the lattice $n_f=2$ result \cite{Ayala:2012pb} which simulates 2 light dynamical quarks, and so this is an apparent success of their approach. For our purposes the main point is that (\ref{e2}) provides a determination of the unquenched gluon mass function when unquenched lattice results for the propagators are used. Note that the SD equation as derived, being homogeneous in $m^2(q^2)$, is blind to operator mixing between $G_{\alpha\beta}G^{\alpha\beta}$ and $m_\psi\overline{\psi}\psi$ and so its results can only apply to massless QCD, $m_\psi=0$.

Given that our interest is in the asymptotic behavior of $m^2(q^2)$ we should ensure that the analysis reflects the known asymptotic behavior of QCD as much as possible. In particular since the propagator functions $\Delta(q^2)$ and $D(q^2)$ are input into the SD equation, it is simple to introduce their correct asymptotic behavior. From the renormalization group this is
\begin{equation}
\Delta(q^2)\rightarrow \ln(q^2)^\gamma/q^2,\quad\quad D(q^2)\rightarrow \ln(q^2)^\delta/q^2
\end{equation}
with $\gamma=-(13C_A-4n_f)/(22C_A-4n_f)$ and $\delta=-9C_A/(44C_A-8n_f)$ in Landau gauge. $\gamma=-31/58$ and $\delta=-27/116$ for $n_f=2$ and $C_A=3$ for $SU(3)$.

We shall implement this asymptotic behavior while fitting the propagator functions to the $n_f=2$ lattice results, which exist for a range of momenta up to $q^2\sim\mu^2$. In particular we fit $q^2\Delta(q^2)$ to the SDE curve on the first plot in Fig.~(9) in \cite{Aguilar:2013hoa} and $q^2D(q^2)$ to the green curve on the first plot in Fig.~(4) in \cite{Ayala:2012pb}. We can obtain good fits via the following simple fitting functions, which are only meant to extend down to $\sim10^{-3}$ GeV$^2$.
\begin{eqnarray}
\Delta(q^2)^{-1}&=&m_0^2+q^2[a+b\ln(q^2+c)^{-\gamma}]\\
D(q^2)^{-1}&=&q^2[d+e\ln(q^2+f)^{-\delta}]\label{e3}
\end{eqnarray}
$m_0$ is set to the $q^2=0$ lattice value $m_0=.413$ GeV and
\begin{equation}
a=-.610,\quad b=.885,\quad c=2.83,\quad d=-.358,\quad e=1.08,\quad f=1.18
.\end{equation}

The integral in (\ref{e2}) can be reduced to an angular integral over $\theta$ ($\cos\theta=qk/\sqrt{q^2k^2}$) and an integral over $k^2$ (see eq.~(8.2) in \cite{Binosi:2012sj} for the explicit result). For each $q^2$ and $k^2$ the angular integral can be performed numerically. The integral equation can be discretized in $\log q^2$ and $\log k^2$ and since it is linear in $m^2(k^2)$ it can be converted into a matrix eigenvalue problem. Solutions only exist for discrete values of $\alpha_s$ and we choose the solution with the lowest (positive) value of $\alpha_s$.\footnote{This is not an unphysical tuning since in an asymptotically free theory the coupling increases into the infrared until the solution develops.} We obtain $\alpha_s=1.145$.

In Fig.~(1a) we compare the mass functions $m^2(q^2)$ to the simple form $\hat{m}^2(q^2)=m_0^2\kappa^2/(q^2+\kappa^2)$ where $\kappa$ is adjusted to match the low momentum behavior, $\kappa^2\approx.2$. We see that $m^2(q^2)$ turns briefly negative.\footnote{This behavior also occurs in the quenched case. J.~Papavassiliou (private communication) confirms that their solution is also not positive definite.} The presence of zeros in the solution is related to the fact that the kernel of the SD equation is not positive definite.  A solution which is everywhere positive only exists for a negative value of $\alpha_s$. Fig.~(1b) displays $q^2m^2(q^2)$ which highlights its completely different high momentum behavior when compared to $q^2\hat{m}^2(q^2)$. Not only does $m^2(q^2)$ not fall monotonically to zero, but it approaches zero significantly faster than $1/q^2$.

\begin{figure}[t]
\vspace{-0ex}
\centering\includegraphics[scale=.28]{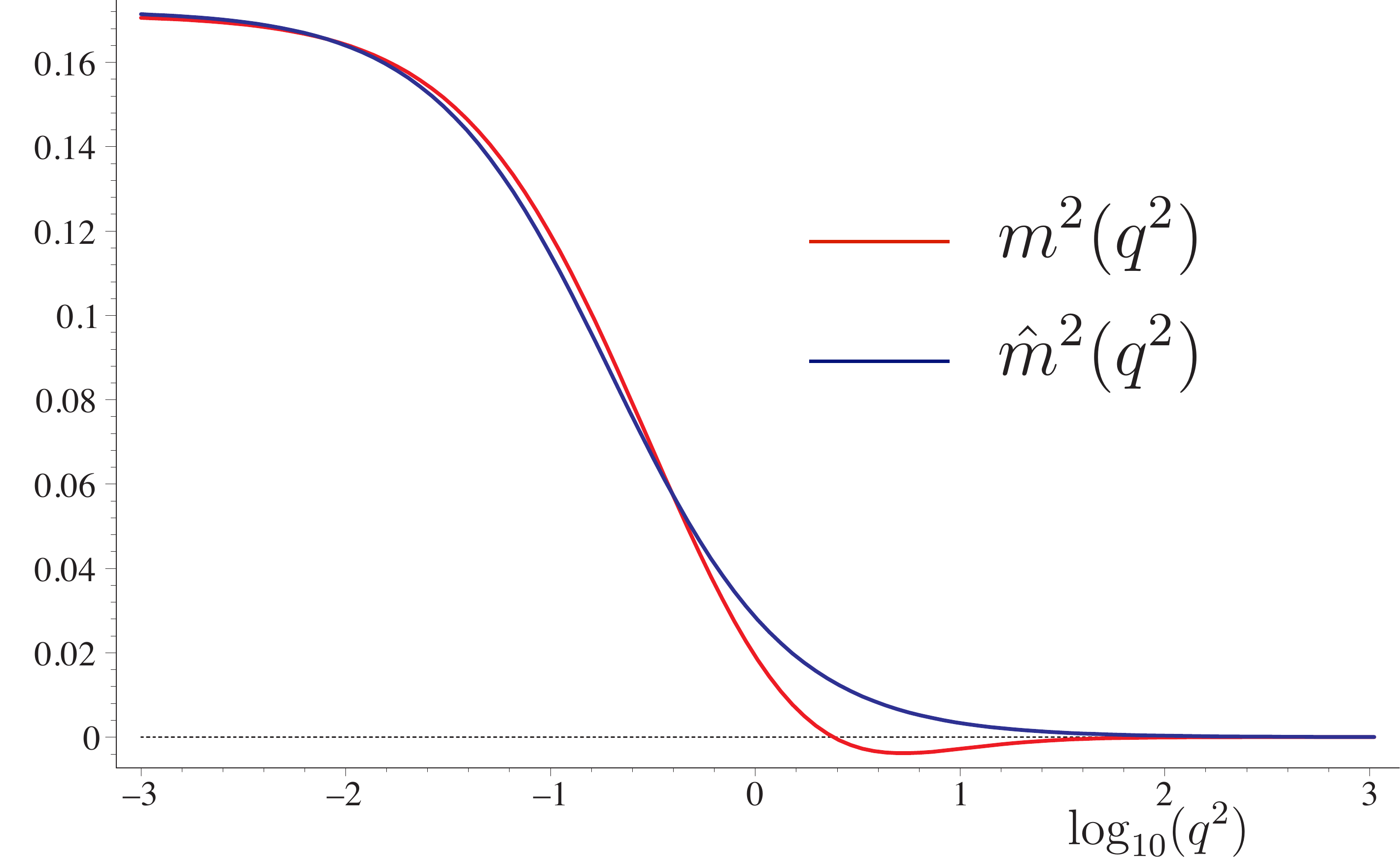}
\centering\includegraphics[scale=.28]{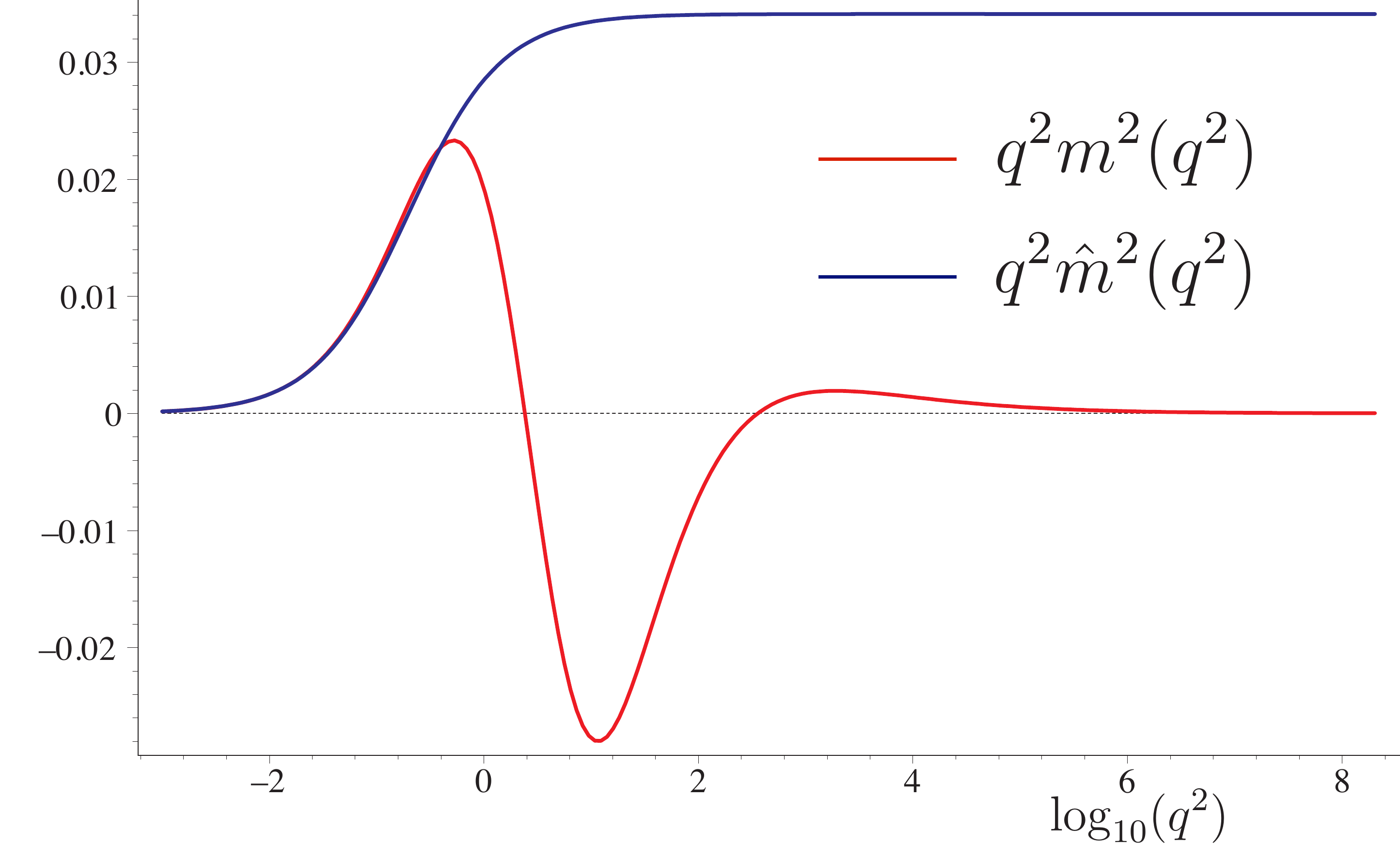}
\vspace{-0ex}
\caption{The gluon mass function $m^2(q^2)$ compared to the simple form $\hat{m}^2(q^2)=m_0^2\kappa^2/(q^2+\kappa^2)$. Note the different ranges of $q^2$.}\end{figure}
\begin{figure}[t]
\vspace{-0ex}
\centering\includegraphics[scale=.28]{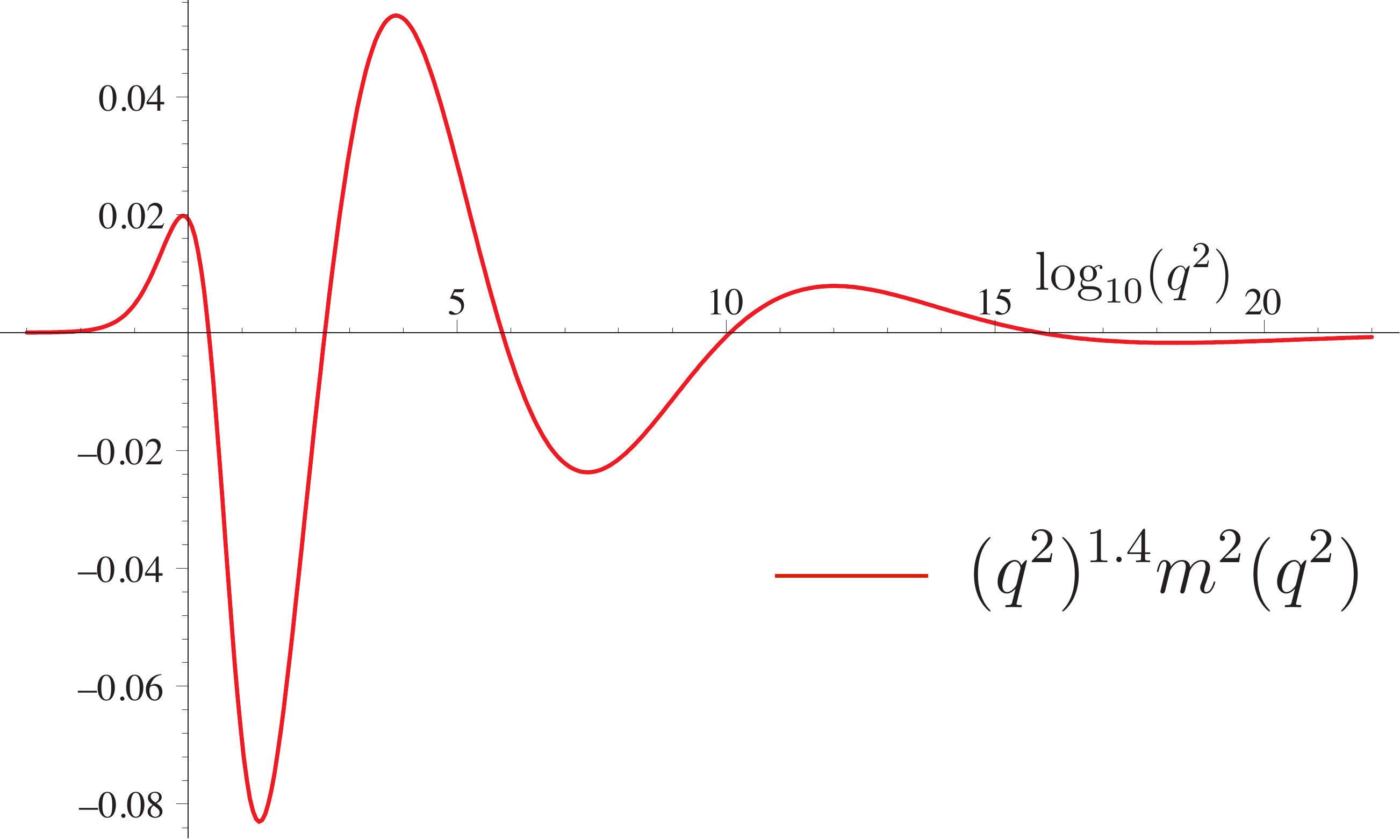}
\centering\includegraphics[scale=.28]{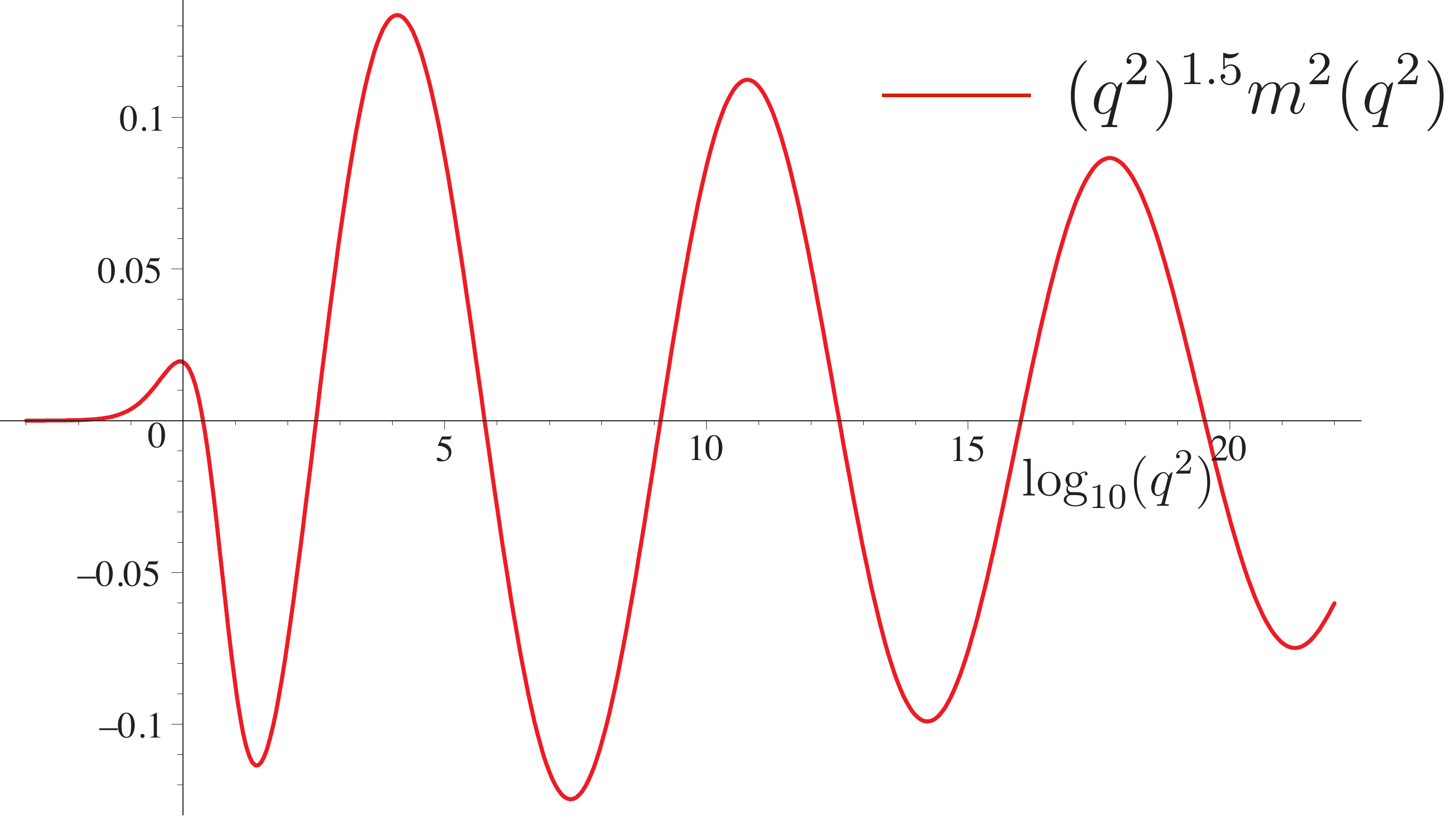}
\vspace{-0ex}
\caption{a) The result of a slightly different fit to the lattice data and b) adding the effect of additional massive quarks as well. Note the increased range of $q^2$.}\end{figure}

In this example $m^2(q^2)$ has two zeros and in fact it is very close to having more. For example we can consider a slightly different fit based on
\begin{equation}
\Delta(q^2)^{-1}=m_0^2+q^2[a+b\ln(q^2+c)\ln(q^2+5^2)^{-\gamma-1}]
\label{e4}\end{equation}
This produces an equally good fit to the lattice data with
\begin{equation}
a=.036,\quad b=.589,\quad c=2.14
,\end{equation}
but the onset of the correct asymptotic behavior is somewhat delayed. The smaller value of $b$ means that this propagator falls slightly slower at large $q^2$ than the previous fit. The change in the resulting gluon mass shows up at very large $q^2$, as seen in Fig.~(2a), where we see that more zeros have developed at high $q^2$. We can also consider the effect that additional massive quarks will have on the anomalous dimensions $\gamma$ and $\delta$. For example introducing the factor $(\log(q^2+m_Q^2)/\log(m_Q^2))^{-36/203}$ in the $b$ term in (\ref{e4}) and the factor $(\log(q^2+m_Q^2)/\log(m_Q^2))^{18/203}$ in the $e$ term in (\ref{e3}) accounts for 4 additional heavy quarks with masses $\sim m_Q$. For $m_Q=100$ GeV this yields Fig.~(2b), showing that we are now firmly in an oscillatory regime. Finally we note that decreasing the $\mu$ in (\ref{e5}) while keeping everything else the same also increases the tendency of the solutions to oscillate.

We may also wonder about the effect of incorporating a running gauge coupling directly into the kernel of the SD equation. This goes beyond the scope of the present SD equation so we only comment on the qualitative effect. We can consider a naive replacement $\alpha_s\rightarrow\alpha_s(\max(k^2,q^2))$ in both the order $\alpha_s$ and $\alpha_s^2$ terms in the kernel, where $\alpha_s(q^2)$ falls logarithmically with $q^2$. The effect is to dampen the oscillations, reduce the number of zeros (typically to two as in Fig.~(1) and not less than one) and to cause $m^2(q^2)$ to approach zero even faster than before. The latter effect is familiar from the quark SD equation; a running coupling causes the mass function to fall more rapidly than a walking coupling.

From these observations our conclusion regarding the asymptotic behavior of $m^2(q^2)$ is that it certainly falls faster than $1/q^2$ and that it is also not described by $1/q^2$ times some inverse power of $\log(q^2)$. From the OPE (\ref{e7}) this is only consistent with a vanishing gluon condensate. We have mentioned that only gauge invariant contributions are displayed in (\ref{e7}) and so one caveat is the possibility that a gauge dependent contribution to the OPE cancels a non-vanishing gluon condensate. But there is no reason for such a cancellation, which in particular would be an accidental cancellation in the Landau gauge chosen for the SD analysis. Another caveat is that the vanishing gluon condensate result might not survive further improvements in the SD analysis. One can keep these caveats in mind, but here we would like to consider some more the meaning of this result $\langle\Theta_\alpha^\alpha\rangle=0$ for massless QCD, should it be true.

SD analyses are often associated with the evaluation of vacuum energies, since effective potentials can be constructed as functionals of the mass functions such that their extrema give back the SD equations. These effective potentials are useful to compare the vacuum energy of different possible extrema. Only relative energies have meaning in this context since a constant can always be added to the effective potential. The loop expansion that defines the effective potential involves massive propagators and is thus a step away from the original theory of the fundamental massless fields. Massive propagators means that an apparent vacuum energy will arise at each order of this loop expansion, unlike the original perturbation theory. But a renormalization scheme can be adopted in this effective description to cancel the finite contributions to the vacuum energy (of order $M^4$ where $M$ is the mass gap) at each order. We are suggesting that this must be done to bring the vacuum energy of the ground state in the effective potential description into line with $\langle\Theta_\alpha^\alpha\rangle=0$.

The SD approach is blind to some known nonperturbative aspects of gauge theories. For example it is blind to instantons, but the instanton contribution to vacuum energy (as reflected in a $\theta$ dependence) vanishes in the presence of massless fermions. The SD approach is also blind to the presence of Gribov copies and the effect that they have on the definition of the path integral \cite{a2}. In fact Gribov copies are associated with a gluonic mass gap and an infrared fixed point. In the Gribov-Zwanziger approach \cite{a2,a3,Dudal:2008sp} these effects are modeled by a modified action containing a new explicit mass parameter.  While the focus in this approach is to model the infrared physics, the effect on the ultraviolet would be such as to violate $\langle\Theta_\alpha^\alpha\rangle=0$. But it has been argued in \cite{Holdom:2009ws,Holdom:2009ma} that it is more realistic to expect that the extremely nonlocal character of Gribov copies should give rise to much softer corrections in the ultraviolet, thus preserving $\langle\Theta_\alpha^\alpha\rangle=0$.

Thus far we have discussed $\langle\Theta_\alpha^\alpha\rangle=0$ as an ultraviolet boundary condition that may be satisfied by massless QCD. Suppose we try to elevate this condition and apply it to the theory of particles and mass more generally, setting aside for now the theory of gravity. First we note that $\langle\Theta_\alpha^\alpha\rangle=0$ prohibits any massive fundamental scalar fields, including the case of the Coleman-Weinberg mechanism, since any such field contributes to vacuum energy. On the other hand $\langle\Theta_\alpha^\alpha\rangle=0$ does not prevent a hierarchy of masses developing below $M$ in a gauge theory of fermions, where $M$ is now just the largest mass gap that develops.

Extended technicolor theories \cite{a4,a5} were just such an attempt to obtain a nontrivial mass spectrum while starting from a massless gauge theory. In these theories $M$ is also a scale at which some gauge symmetry breaks. Another symmetry breaking could occur on a mass scale $m_1\ll M$, with a fermion condensate of order $m_1^3$. Even smaller fermion masses can then appear of order $m_2\sim m_1^3/M^2$. It may only be for $q\gtrsim M$ that the $\sim 1/q^2$ behavior of these latter masses become apparent. To describe this behavior in terms of the OPE would imply  condensates of these fermions of order $m_2M^2$, and this agrees with the one loop estimate using $M$ as the cutoff. The eventual soft asymptotic behavior remains consistent with vanishing explicit masses for all fermions.

The low energy effective theory can contain small Lagrangian level masses of the type $m_2$. If these are the quarks of QCD then an apparent gluon condensate will be induced by these quark masses, due to the operator mixing mentioned above. But the $m_2\overline{\psi}\psi$ mass term only exists in the low energy description and so the gluon mass function will eventually fall as found above for sufficiently large momentum. The OPE still points to a vanishing gluon condensate. Similarly the effective theory leads to apparent contributions to vacuum energy, at the very least of size $m_2^4\log(M/m_2)$. Such finite contributions should again be subtracted at each order in perturbation theory for consistency with $\langle\Theta_\alpha^\alpha\rangle=0$.

In summary we have used the operator production expansion to relate the gluon condensate to the asymptotic behavior of the gluon mass function. We have found that the extraction of this asymptotic behavior from a recently developed Schwinger-Dyson equation for the gluon mass points to a vanishing gluon condensate. From the trace anomaly relation this also points to a vanishing vacuum energy in massless QCD. This approach shows that a vanishing vacuum energy is a statement about the softness of the ultraviolet behavior of the fundamental fields. The SD analysis also shows how this result for an asymptotically free massless gauge theory can be consistent with a mass gap. This consistency may be more firmly established as the SD analysis in the gluon sector continues to improve. For example we have argued that the incorporation of the running coupling into the SD kernel will only strengthen the result.

It is also possible that lattice studies can address the question of the gluon condensate in massless QCD more directly, rather than just providing input to the SD equation. We encourage more lattice studies along the lines of \cite{D'Elia:1997ne,Meggiolaro:1998yn}, where the behavior of a field strength correlator was used to extract the gluon condensate (which in that study was consistent with a vanishing value in the chiral limit). The main message is that one should not take for granted the existence of a vacuum energy in a massless gauge theory of fermions. And since there remains a possibility that a massless gauge theory underlies all of particle physics, there appears to be ample motivation for further studies.

\section*{Acknowledgments}
This work was supported in part by the Natural Science and Engineering Research Council of Canada.

\end{document}